\documentclass[twocolumn,preprintnumbers,amsmath,amssymb]{revtex4}
\usepackage{epsfig}
\begin{document}

\title{Finite size effects on spin-torque driven ferromagnetic resonance in spin-valves with a Co/Ni synthetic free layer}

\author{W. Chen, G. de Loubens, J-M. L. Beaujour, A. D. Kent}
\affiliation{Department of Physics, New York University, New York,
NY 10003}
\author{J. Z. Sun}
\affiliation{IBM T. J. Watson Research Center, Yorktown Heights, NY
10598}
\date{October 18th, 2007}

\begin{abstract}
Spin-torque driven ferromagnetic resonance (ST-FMR) is used to study
magnetic excitations in Co/Ni synthetic layers confined in
nanojunctions. Field swept ST-FMR measurements were conducted with a
magnetic field applied perpendicular to the layer surface. The
resonance lines were measured under low amplitude excitation in a
linear response regime. The resulting resonance fields were compared
with those obtained using conventional rf field driven FMR on
extended films with the same Co/Ni layer structure. A lower
resonance field is found in confined structures. The effect of both
dipolar fields acting on the Co/Ni layer emanating from other
magnetic layers in the device and finite size effects on the spin
wave spectrum are discussed.
\end{abstract}
\maketitle

One approach to study ferromagnetic resonance (FMR) of a magnetic
layer in a confined structure is to use the spin transfer
interaction \cite{Slonczewski1996, Berger1996} in a
current-perpendicular (CPP) nanojunction. An rf current is applied
to a magnetic tunnel junction \cite{Tulapurkar2005} or spin valve
\cite{Sankey2006}, to drive FMR, in a method known as the
spin-torque-driven ferromagnetic resonance (ST-FMR). This new
technique enables quantitative studies of magnetic properties of
materials in nanopillars, such as their magnetic excitations,
anisotropy and damping.

Spin-transfer devices that incorporate materials with perpendicular
magnetic anisotropy are of great interest. This is because of their
potential to lead to faster ST-devices, with lower power dissipation
\cite{Kent2004} and critical current \cite{Sun2000}. Recently,
Mangin \emph{et al.} studied perpendicular spin valves with a Co/Ni
multilayer free layer, where a large magnetoresistance value and a
high spin torque efficiency were observed \cite{Fullerton2006}.

In this work, we present ST-FMR studies on bilayer spin valves,
where the thin (free) layer is composed of a Co/Ni synthetic layer
and the thick (fixed) layer is pure Co. By comparing the ST-FMR
resonance fields with those of conventional rf field driven FMR of
extended Co/Ni films with the same layer structure, we illustrate
interactions important in ST-FMR of nanojunctions. Specifically, we
discuss both dipolar interactions between the Co/Ni layer and other
magnetic layers in the device, and finite size effects on the
magnetic excitation spectrum.

\begin{figure}[t]
\includegraphics[width=0.20\textwidth, angle=270]{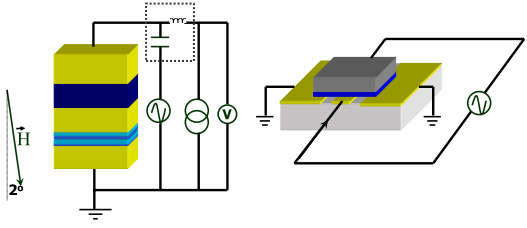}
\caption{(a): Spin valve layer structure and ST-FMR circuit. (b):
Field-driven FMR on same-stack extended films using the flip-chip
method.}\label{Layout}
\end{figure}

Pillar junctions with submicron lateral dimensions and rectangular
shape, shown in Fig. \ref{Layout}a, were patterned on a silicon
wafer using a nanostencil process \cite{Sun2002}. Junctions were
deposited by evaporation, and have the layer structure
$\parallel1.5$ nm Cr$\mid 100$ nm Cu$\mid 20$ nm Pt$\mid 10$ nm
Cu$\mid$ [0.4 nm Co$\mid$ 0.8 nm Ni]$\times 3 \mid 10$ nm Cu$\mid
12$ nm Co$\mid 200$ nm Cu$\parallel$. The ST-FMR measurement setup
is shown in Fig. \ref{Layout}(a). An rf current generated by a high
frequency source is coupled with a dc current through a bias-T (the
dashed-line box in Fig. \ref{Layout}(a)) into the spin valve. At
resonance, the rf current and spin valve resistance oscillate at the
same frequency. This results in a dc voltage ($V=<I(t)R(t)>$)
\cite{Tulapurkar2005, Sankey2006}. Assuming a small angle circular
precession of the free layer on resonance,
\begin{eqnarray}
  {V={{1}\over{4}}(R_{AP}-R_{P})I_{\text{rf}}\sin \beta \sin
\theta} \label{eq1}
\end{eqnarray}
where $\beta$ is the angle between the free and fixed layers (before
applying the rf current) and $\theta$ is the precession angle.
$I_\text{rf}$ represents the rf current amplitude in the junction,
and $R_{AP}$($R_{P}$) is the static junction resistance when free
layer and fixed layer are antiparallel (parallel) to each other.
With a perpendicular magnetic field greater than the free layer's
easy-plane anisotropy field, the free layer magnetization is normal
to the surface, while the fixed layer, which has a larger easy-plane
anisotropy field, is still mainly magnetized in the film plane. In
this way, the signal is maximized, according to Eq. \ref{eq1}. To
improve the signal (typically in the sub-$\mu$V range) to noise
ratio, we modulate the rf current on and off at 800 Hz and use a
lock-in amplifier to detect the voltage at this frequency.

Extended films with the same stack as the free layer were deposited
on an oxidized Si wafer using the same deposition technique. It has
the same Co/Ni synthetic layer sandwiched between 10 nm of Cu on
each side. As is shown in Fig. \ref{Layout}(b), it was measured by
placing the film onto a 50-Ohm-matched coplanar waveguide
\cite{Beaujour2006}. An rf current was sent through the waveguide
and generates an rf field that drives the magnetic film into
resonance. The transmission of the rf signal was measured using a
network analyzer, as a function of rf frequency and external
magnetic field.

\begin{figure}[t]
\includegraphics[width=0.30\textwidth, angle=270]{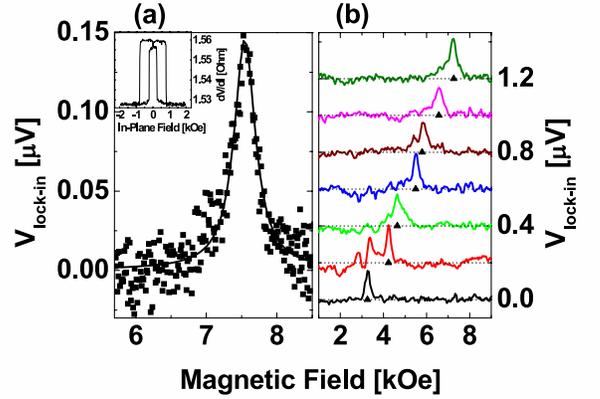}
\caption{(a): ST-FMR voltage signal ($\blacksquare$ points) as a
function of applied perpendicular magnetic field together with a
Lorentzian fit (solid line). The measurement was done on a
50$\times$150 nm$^{2}$ junction with an rf amplitude of
$I_\text{rf}$=560 $\mu$A at a frequency of 18 GHz. Inset: MR
hysteresis loop on the same junction with the magnetic field applied
in-plane. (b): Zero dc current lock-in voltage signal as the
function of applied magnetic field at different frequencies from 4
GHz up to 16 GHz in 2 GHz steps.}\label{Resonance}
\end{figure}

The magnetoresistance (MR) of the nanojunctions was measured using a
four-point geometry with the magnetic field applied in the film
plane. A typical MR hysteresis loop of a 50$\times$150 nm$^{2}$
junction is shown in the inset of Fig. \ref{Resonance}(a).
MR$=(R_{AP}-R_{P})/R_{P}$, is $\simeq ~2.3 \pm 0.2\%$ for a total of
10 junctions studied.

ST-FMR measurements were conducted with the external magnetic field
$H_\text{app}$ applied \textit{nearly} perpendicular to the film
plane ($2^{\circ}$ off the normal direction as shown in Fig.
\ref{Layout}(a), where the small in-plane component is along the
easy-axis of the junction, in order to avoid vortex states in the
free layer). This was measured using a two-point geometry. Fig.
\ref{Resonance}(a) shows a typical field-swept resonance line at a
fixed rf frequency of 18 GHz and zero dc current. It was measured on
the same junction on which the data in the inset was taken. The
resonance is fit by a Lorentzian, indicated by the solid line. From
the peak height $V_\text{peak}$, we estimate the precession angle to
be $1.9^{\circ}$ using Eq. \ref{eq1}. We verified that this data was
taken in a linear response regime with $V_\text{peak}/I_\text{rf}^2$
independent of $I_\text{rf}$.

A series of ST-FMR resonance lines at different rf frequencies $f$
were measured within the low amplitude linear regime. Those with 7
different rf frequencies (4$\sim$16 GHz in 2 GHz steps) are plotted
in Fig. \ref{Resonance}(b), with each adjacent curve offset by $0.2
\: \mu$V. The resonance field $H_\text{res}$ ($\blacktriangle$ in
Fig. \ref{Resonance}(b)) increases linearly with $f$ greater than 4
GHz. At $f<$4 GHz, the perpendicular magnetic field at resonance is
lower than the easy-plane anisotropy field of the free layer.
Therefore, the free layer magnetization tends to tilt into the
plane, leading to a lower resonance field. Similar dispersion
relationships have been found on junctions with other lateral
dimensions.

\begin{figure}[t]
\includegraphics[width=0.30\textwidth, angle=270]{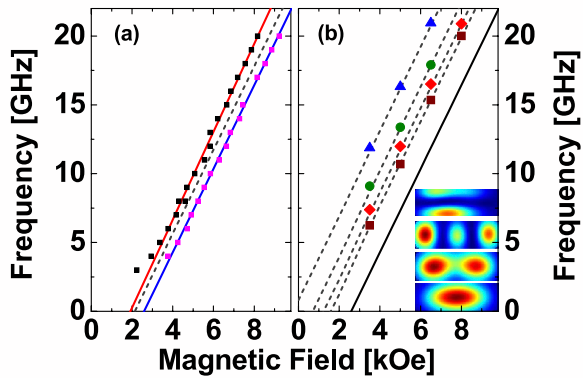}
\caption{(a): Comparison of resonance field as the function of
frequency between the spin valve junction (black dots) and
same-stack extended film (pink dots). The spin valve junction is a
different one from that in Fig. 2, but has the same lateral
dimension. Solid lines are linear fits. Dashed line: linear fit of
resonance field with estimated dipolar fields corrected. (b): The
dispersion of the lowest four spin wave modes on a
50$\times$150$\times$3.6 nm$^{3}$ Co/Ni synthetic structure using
OOMMF simulation (dots) and the analytical model discussed in the
text (dashed lines). Solid line is the linear fit of $H_\text{res}$
of the extended film. Corresponding mode profiles are shown in the
lower-right corner, in the order of ($n_{x}, n_{y}$)=(1, 1) (1, 2)
(1, 3) (2, 1) from bottom to top.}\label{Comparison}
\end{figure}

The resonance field for both extended films and nanojunctions as the
function of $f$ are plotted in Fig. \ref{Comparison}(a).  Black dots
are for a similar spin valve with the same lateral dimension and
pink dots for the same-stack extended film. Further details on the
conventional FMR experiments can be found in Ref.
\cite{Beaujour2007}. Red and blue solid lines are their
corresponding linear fits above 4 GHz. By comparing these two sets
of data, we find a slight difference in slope and a small shift in
the zero frequency intercept. The relationship between $f$ and
$H_{\text{res}}$ in the extended film is given as:
${{h}\over{\mu_{B}}}f=\text{g}(H_\text{res}-4\pi M_\text{eff})$
\cite{Kittel} in the case where the magnetization is normal to the
film surface. Here g is the Land\'{e} g factor of the film, and the
easy-plane anisotropy is $4\pi M_{\text{eff}}=4\pi
M_{\text{s}}-H_\text{P}$, where $M_{\text{s}}$ and $H_\text{P}$
represent the saturation magnetization, and the perpendicular
anisotropy field. A direct linear fit of each data set gives a slope
g=2.17, field-axis intercept $4\pi M_\text{eff}=2.58$ kOe for the
extended film, and a slightly larger slope (2.28) and a smaller
field-axis intercept (1.92 kOe) for the confined structure in the
spin valves. Such a consistency confirms that the main peak of the
ST-FMR signal comes from the Co/Ni synthetic free layer rather than
the fixed Co layer \cite{Footnote1}.

We now estimate the effect of dipolar fields $H_\text{dip}$ from
other magnetic layers in spin valves, which come from the fixed Co
layer and the junction level magnetic residuals outside the stencil
holes \cite{Sun2002}. $H_\text{dip}$ from the normal component of
the fixed Co layer is not negligible. At $f=10$ GHz, the Co layer is
tilted $\sim16^{\circ}$ out-of-plane at $H_\text{res}$, and the
component of $H_\text{dip}$ normal to the film surface is 500 Oe. At
higher frequencies, the normal component of $H_\text{dip}$ is larger
since the fixed Co layer magnetization is more tilted out-of-plane
for larger $H_\text{res}$. $H_\text{dip}$ of the junction level Co
residual is estimated to be $\sim$15 times smaller, while that of
the junction level Co/Ni residual has a constant -120 Oe field
contribution normal to the surface. The in-plane component of the
dipolar fields ($\lesssim$ 260 Oe) shifts $H_\text{res}$ down by
$\sim$30 Oe at 4 GHz, and this shift is significantly reduced at
higher frequencies. Therefore, the dipolar fields from the normal
component of the fixed Co layer and from the junction level Co/Ni
residual are important in biasing $H_\text{res}$ of free layer in
spin valves. The dashed line in Fig. \ref{Comparison}(a) is the
linear fit of $H_\text{res}$ with the dipolar field correction. In
the new linear fit, the slope becomes 2.18, which is almost the same
as the g factor of the extended film, and the field-axis intercept
becomes 2.13 kOe. A 450 Oe shift between them at all frequencies
still remains.

Similar results have been found in other spin valve junctions.
ST-FMR results taking into account the dipolar fields from other
magnetic layers in the device show a similar g-factor
(2.18$\pm$0.03) with a 370$\sim$570 Oe lower resonance field than
that of the extended film. Numerical calculation on the normal spin
wave (SW) modes in elliptical Py disks was presented in Ref.
\cite{McMichael2005}, and the resonance shift of different SW modes
due to the finite size effect was discussed. In order to estimate
the resonance shift in our rectangular Co/Ni synthetic nanoelement,
we have also done micromagnetic simulations using OOMMF
\cite{OOMMF}, which includes the Zeeman, exchange and magnetostatic
contributions to the energy (but not the spin transfer interaction).
The modes of a 50$\times$150 nm$^{2}$ element are shown in
lower-right of Fig. \ref{Comparison}(b). The mode resonance is
shifted towards lower field as the order of the SW modes becomes
higher (dots in Fig. \ref{Comparison}(b)). The lowest order SW mode
is shifted by $\sim$1 kOe from the thin film result (solid line).
This shift is the same order of magnitude as found in our
experiments. It is larger than that observed possibly because the
lateral dimensions of the pillar are larger than the nominal
dimensions. The 2nd mode is close to the lowest one, while 3rd and
4th ones are further apart. This shows the effect of finite size and
mode structure on the resonance field. Furthermore, multiple peaks
found at several frequencies (see Fig. \ref{Resonance}(b) at $f=6$
GHz, for example), may be due to the excitation of higher order SW
modes.

An analytical estimation of the energy of the SW modes in a
rectangular element enables a better understanding of the physics.
The idea is to assume a sinusoidal profile of the eigenmode in the
finite rectangular element \cite{Bailleul2006}, and to use the
dipole-exchange dispersion of the SW modes in a perpendicularly
magnetized film \cite{Slavin1986}:
\begin{eqnarray}
  \omega_{k}^{2}=\gamma^{2}(H_\text{in}+{{2A k^{2}}\over{M_\text{s}}})[H_\text{in}+{{2A k^{2}}\over{M_\text{s}}}+4\pi(1-{{1-e^{-kt}}\over{kt}})M_\text{s}]\label{eq2}
\end{eqnarray}
where $\gamma$ is the gyromagnetic ratio, and the internal field
$H_\text{in}=H_\text{app}-4\pi N_{z}M_{\text{s}}+H_\text{P}$, where
$N_{z}$ is the demagnetization factor in the normal direction. ${2A
k^{2}}\over{M_\text{s}}$ is the exchange term, with A and k
representing exchange constant and in-plane wave vector. $4\pi
(1-{{1-e^{-kt}}\over{kt}})M_\text{s}$ is the dipolar term in which t
is the thickness of the disk, and it describes the SW dynamic
dipole-dipole interaction. Moreover, as discussed in Ref.
\cite{Guslienko2002}, the oscillating magnetization is only
partially pinned at the boundary. Thus, a larger effective lateral
dimension needs to be introduced in the analytical calculation to
mimic the partial pinning. The 1 kOe shift, like that in the OOMMF
simulation, can be found using Eq. \ref{eq2} by introducing an
effective lateral dimension $L_{x}\times L_{y}$ $\sim$70$\times$170
nm$^{2}$. Using these dimensions, the higher order modes denoted as
($n_{x}$,$n_{y}$) are also quantitatively reproduced with
$k_{x,~y}=n_{x,~y}\pi/L_{x,~y}$, $n_{x,~y}\in \mathbb{N} ^{*}$ (see
dashed lines in Fig. \ref{Comparison}(b)).

In summary, we used the ST-FMR technique to measure the resonance
fields of Co/Ni synthetic layers in a confined spin valve structure
and compared it to that of the same-stack extended film. The effects
of dipolar fields and finite element size produce changes in the
resonance field that are the same order of magnitude as those
observed in the experiment.

This research is supported by NSF-DMR-0706322, ARO-W911NF-07-1-0643
and an NYU-Research Challenge Fund award.


\begin{thebibliography}{17}
\bibitem{Footnote1} FMR measurement on 10 nm thick extended pure Co layers shows that the resonance field is much larger (a few Tesla) for $f>$3 GHz, because of its high
easy-plane anisotropy (17 kOe). At lower frequencies, the resonance
field decreases rapidly as the layer magnetization tilts into the
plane, with a slope far from the g$\simeq2$ seen in the ST-FMR
studies.
\end{thebibliography}
\end{document}